\documentclass[11pt,fleqn]{article}
\usepackage{amsmath,amssymb,graphicx,epsfig}
\usepackage{cancel}
\numberwithin{equation}{section}
\usepackage{color}

\font\tenrsfs=rsfs10
\font\sevenrsfs=rsfs7
\font\fiversfs=rsfs5
\newfam\rsfsfam
\textfont\rsfsfam=\tenrsfs
\scriptfont\rsfsfam=\sevenrsfs
\scriptscriptfont\rsfsfam=\fiversfs

\title{Gauge R-symmetry and de Sitter vacua in supergravity and string theory}

\begin{document}

\setlength{\baselineskip}{7mm}
\begin{titlepage}

\begin{flushright}

{\tt CERN-PH-TH/2014-034}\\

\end{flushright}

\vspace{1cm}
\begin{center}

{\huge \bf  Gauged R-symmetry and de Sitter vacua in supergravity and string theory

\vspace{0,3cm}

} 

\vspace{1cm}
{ \large{ \bf{ Ignatios Antoniadis}} $^{\star,}$\footnote{${}$ On leave of absence
from CPHT {\'E}cole Polytechnique, F-91128, Palaiseau Cedex,France.}
\large{ \bf{and Rob Knoops}}$^{\star,+}$

\bigskip
\centerline{${}^\star $  Department of Physics, CERN Theory Division} 
\centerline{ CH-1211 Geneva 23, Switzerland}
\bigskip
\centerline{${}^+$  Instituut voor Theoretische Fysica, KU Leuven}
\centerline{Celestijnenlaan 200D, B-3001 Leuven, Belgium}
\bigskip
}

\end{center}
\vspace{25pt}

\centerline{{\bf Abstract}}

\vspace{10pt}

\noindent

 A new class of metastable de Sitter vacua is presented and analyzed in $\mathcal N =1$ supergravity and string theory with tunable (infinitesimally small) value of the cosmological constant. They are based on a gauged R-symmetry and the minimal spectrum contains a vector and a linear multiplet of the string dilaton or of a compactification modulus. In the minimum of the scalar potential supersymmetry is broken by an expectation value of both a D- and F-term. 

\begin{center}
\end{center}

\vspace{150 pt}


\end{titlepage}

\newpage

\pagestyle{plain}

\renewcommand{\theequation}{\thesection.\arabic{equation}}

\setcounter{page}{1}

\section{Introduction}
de Sitter (dS) vacua in $ \mathcal N=1$ supergravity and string theory with tunable (infinitesimally small) value of the cosmological constant are rare and their study is therefore interesting and challenging \cite{Maloney:2002rr}. 
In this work, we construct and analyze a new class of such metastable solutions based on a gauged R-symmetry \cite{Freedman:1976uk} (For other studies based on a gauged R-symmetry, see for example \cite{Cremmer:1983yr}). 
The minimal field content, besides the gravity multiplet, consists of one vector multiplet and one chiral multiplet on which the R-transformation (up to an appropriate field redefinition) 
acts as a shift along the imaginary part of its scalar component~\cite{Villadoro:2005yq}. 
In the rigid (globally supersymmetric) limit, the vector multiplet decouples and one is left with a single chiral multiplet $S$ and a shift symmetry. Obviously, the most general Kahler potential is a function of the real part $S+{\bar S}$ while the only allowed superpotential is a single exponential $e^{bS}$ or a field independent constant. In the latter case the shift transformation is not part of an R-symmetry.

In the context of string theory $S$ is a compactification modulus or the universal dilaton and the exponential superpotential may be generated by non-perturbative effects, either in the string world-sheet or in space-time (such as gaugino condensation), respectively. In the following, we consider as example the case of the string dilaton although our expressions can be generalized to a generic modulus in a straightforward way. A constant superpotential on the other hand corresponds to a tree-level contribution obtained when string theory is considered away from its critical dimension and is proportional to the central charge deficit~\cite{Antoniadis:1990uu, Antoniadis:2008uk}.

The presence of the shift symmetry implies that there is an alternative formulation in terms of a linear multiplet containing a 2-index antisymmetric tensor potential which is dual to ${\rm Im}S$ (under Poincar\'e duality of its field-strength). 
The 2-form field appears naturally in the string basis and, as we show, the linear multiplet description of the theory persists even in the presence of the superpotential. 
We therefore obtain a theory, dual to the one described in~\cite{Villadoro:2005yq}, where the minimal field content contains, besides the gravity multiplet, a vector multiplet and a real multiplet whose propagating degrees of freedom coincide with the fields of a linear multiplet.

The resulting scalar potential is runaway but can be stabilized in the presence of a D-term contribution with a Fayet-Iliopoulos (FI) term when the shift symmetry is gauged. Such a contribution can arise from a non-supersymmetric D-brane configuration and is proportional to the total D-brane tension deficit. The case of a non R-symmetry with a constant superpotential was studied in~\cite{Antoniadis:2008uk} and gives rise to an anti-de Sitter (AdS) minimum with broken supersymmetry by a D-term.

In this work we study the case of a gauged R-symmetry with an exponential superpotential quantifying the results of~\cite{Villadoro:2005yq}. 
We show that the scalar potential has a dS metastable minimum with broken supersymmetry by both an F- and D-term and compute the physical spectrum.
We also derive the linear multiplet description where in the supersymmetry breaking minimum the linear multiplet combines with the massless vector multiplet, leading to massive fields: a vector, a scalar and a Majorana fermion (besides of course the graviton and the massive gravitino).

The outline of the paper is the following. In Section 2, we introduce the field content and the notation. In Section 3, we review the linear-chiral multiplet duality in the rigid case without superpotential, while in Section 4, we generalize it in the presence of a superpotential. In Section 5, we extend the analysis in supergravity. In Section 6, we derive the scalar potential, while in Section 7, we perform the minimization and derive the spectrum in the dS metastable minimum. Finally, in the first three appendices we derive the expressions of the Lagrangian in components after the elimination of the auxiliary fields, both in the rigid case (Appendix \ref{dofL}) and in supergravity the bosonic part (Appendices \ref{appendixsugra} and \ref{Proof}). In Appendix \ref{masses} we calculate the masses of the physical particles.

\section{The dilaton and shift symmetry}

In supersymmetric theories the string dilaton and the axion $a$ can be described as the real and imaginary part of the scalar component $s$ of a chiral multiplet $S = (s,\psi,F)$.
\begin{align}
 s = \frac{1}{g^2} + ia, \label{sgaugecoupling}
\end{align}
where $g$ is the four dimensional gauge coupling. In perturbation theory, the axion has an invariance under a Peccei-Quinn symmetry which shifts $s$ by an imaginary constant, 
\begin{align}
s \longrightarrow s - i\alpha, \label{shift}
\end{align}
where $\alpha$ is real. It is well known that there exists a dual description where the axion is described by an anti-symmetric tensor $b_{\mu \nu}$, which has a gauge symmetry given by
\begin{align}
b_{\mu \nu} \longrightarrow b_{\mu \nu}+ \partial_\mu b_\nu - \partial_\nu b_\mu. \label{bgauge}
\end{align}
 In the context of string theory, it is this anti-symmetric tensor which appears in the massless spectrum as the supersymmetric partner of the dilaton.
The antysymmetric tensor is related to a field strength $v_\mu$ by  $v^\mu = \frac{1}{\sqrt 6} \epsilon^{\mu \nu \rho \sigma} \partial_\nu b_{\rho \sigma} $. One can then directly see that $v_\mu$ is divergenceless 
\begin{align} \partial^\mu v_\mu = 0 \label{vdiv},\end{align}
which is the  Bianchi identity.
Together with a real scalar $l$ and a Majorana fermion $\chi$, this real vector $v_\mu$ belongs to a linear multiplet $L$, defined by
\begin{align}
 D^2 L = \bar D^2 L = 0, \label{linearmultiplet}
\end{align}
where the  chiral and anti-chiral projection operators $\bar D^2 = \bar D_{\dot \alpha} \bar D^{\dot \alpha}$ and $D^2 = D^\alpha D_\alpha$ are defined as
\begin{align}
D_\alpha &= \frac{\partial}{\partial \theta^\alpha} + i \sigma_{\alpha \dot \alpha}^{\mu} \bar \theta^{\dot \alpha} \partial_\mu ,\\
\bar D_{\dot \alpha} &= - \frac{\partial}{\partial \bar \theta^{\dot \alpha}} - i \theta^\alpha \sigma_{\alpha \dot \alpha}^{\mu} \partial_\mu \label{chiralprojection},
\end{align}
where $\theta^\alpha (\bar \theta^{\dot \alpha})$ are the fermionic coordinates of the (anti-)chiral superspace.
It should be noted that the shift symmetry (\ref{shift}) of the axionic partner of the dilaton is a crucial ingredient for the duality to work, since it is related to the gauge symmetry (\ref{bgauge}) which in turn is responsible for (\ref{vdiv}).

In order for the theory to have a shift symmetry, the K\"ahler potential must be a function of the combination $S + \bar S$.
Since we are interested in string inspired actions, we will assume a K\"ahler potential of the form
\begin{align}
\mathcal K (S + \bar S) = - p \ln (S + \bar S), \label{Kahlerpot}
\end{align}
where $p$ is a real constant. 

\section{Easy warm-up: No superpotential}
As a warm-up, we will show how one can describe the theory
\begin{align}
\mathcal L = \int d^2 \theta d^2 \bar \theta \mathcal K (S + \bar S),
\end{align}
with $K(S + \bar S)$ given by (\ref{Kahlerpot}), in terms of a linear multiplet. For this, we consider the Lagrangian
\begin{align}
\mathcal L = \int d^2 \theta d^2 \bar \theta \left( \mathcal F(L) - p L(S + \bar S) \right) \label{L_FO_1}
\end{align}
where $L$ is a priori an unconstraint real vector superfield. For purposes that become clear in an instant, we take $\mathcal F(L)$ to be given by $\mathcal F(L) =p \ln(L)$.
The equations of motion for the (anti-)chiral superfields $S$($\bar S$) give $D^2 L = \bar D^2 L = 0$, which is exactly the condition (\ref{linearmultiplet}) for $L$ to be a linear multiplet. 
The second term in (\ref{L_FO_1}) is thus a generalised Lagrange multiplier that ensures the linearity of $L$.
The resulting theory in terms of the linear multiplet is given by
\begin{align}
\int d^2 \theta d^2 \bar \theta \mathcal F(L).
\end{align}
On the other hand, the equations of motion of (the now unconstrained real superfield) $L$ give
\begin{align}
\mathcal F'(L) = p \left( S + \bar S \right),
\end{align} 
so that  
\begin{align} S + \bar S = \frac{1}{L} \label{SdualL}. \end{align}
Some basic algebra then shows that
\begin{align}
\mathcal F(L) =p \ln \left( \frac{1}{S + \bar S} \right) = - p \ln \left(S + \bar S \right).
\end{align}
The chiral theory is then given by
\begin{align}
\int d^2 \theta d^2 \bar \theta \mathcal K(S + \bar S),
\end{align}
with $\mathcal K(S + \bar S)$ given in (\ref{Kahlerpot}).

Since the natural partner of the graviton in the massless sector of superstrings is the scalar dilaton and an antisymmetric tensor, the above exercise should be repeated in supergravity (local supersymmetry). This is done below in section \ref{sugraduality}.

\section{With a superpotential} \label{rigidsusy}
It is now a natural question to ask whether one can add a superpotential to this theory, so that the shift symmetry survives. Since the superpotential $W(S)$ is a holomorphic function, it can not be a function of the combination $S + \bar S$ and thus any (non-constant) superpotential will transform under the shift symmetry. The only possibility is then to cancel this shift by an appropriate R-transformation. It turns out that an exponential superpotential
\begin{align}
W(S) = a \exp (bS) \label{superpotential}
\end{align}
has this property: under (\ref{shift}) the superpotential transforms as $W(S) \longrightarrow W(S) e^{-i\alpha b}$, which can be compensated by the R-transformation $\theta \longrightarrow \theta e^{-\frac{i}{2}\alpha b} $ so that $\int d^2 \theta W(S)$ is invariant under the combined shift and R-transformations. 
Note that one can not gauge the shift symmetry in this model, since it is entangled with an R-symmetry, which can not be gauged in rigid supersymmetry.

We are now interested in the dual description of the theory
\begin{align}
\mathcal L= \int d^4 \theta \mathcal K(S +\bar S) + \int d^2 \theta W(S) + \int d^2 \bar \theta \bar W(\bar S) \label{L_chiral}
\end{align}
with K\"ahler potential and superpotential given by (\ref{Kahlerpot}) and (\ref{superpotential}). 
It is interesting to note that these superpotentials appear naturally in the context of gaugino condensation (see for example \cite{Dine:1985rz}) and in recent works on supersymmetric extensions of Starobinsky models of inflation \cite{Ferrara:2013rsa}.
In the following, we use the conventions of \cite{Wess:1992cp}. 
To find the dual theory, we start from the Lagrangian
\begin{align}
\mathcal L & = \int d^4 \theta \mathcal F(L)  -p \int d^4 \theta L (S + \bar S) + \int d^2 \theta W(S) + \int d^2 \bar \theta \bar W(\bar S) \label{FO_shift},
\end{align}
where as before, $S$ and $\bar S$ are chiral and anti-chiral superfields, respectively, and $L$ is an a priori unconstrained real superfield. As we will see below, it turns out that in the dual theory, L will still be a real superfield off-shell. However, it will contain the same number of propagating degrees of freedom as those of a linear multiplet.
$W(S)$ is given in (\ref{superpotential}) and $\mathcal F(L) = p \ln(L)$.
The equations of motion for $L$ give, as before
\begin{align} S + \bar S = \frac{1}{L} \label{SdualL2}. \end{align}
By substituting this relation into the Lagrangian (\ref{FO_shift}) one retrieves (\ref{L_chiral}).

We now derive the equations of motion for the chiral superfield. This results in modified linearity conditions
\begin{align}
p \bar D^2 L &= -4 W'(S), \label{LW} \\
p D^2 L &= -4 \bar W'( \bar S), \label{LbarW}
\end{align}
and thus $L$ is still an unconstrained real multiplet.
One can use these equations to write $S$ and $\bar S$ in terms of $L$
\begin{align}
S &= \frac{1}{b} \ln \left( \frac{-p}{4ab} \bar D^2 L  \right)\label{SL},  \\
\bar S &= \frac{1}{b} \ln \left( \frac{-p}{4ab} D^2 L \right) \label{SbarL}.
\end{align}
Note that one can rewrite the Lagrangian (\ref{FO_shift}) as
\begin{align}
\mathcal L &=  \int d^4 \theta  \mathcal F(L) + \int d^2 \theta \left( \frac{p}{4} \bar D^2 L S + W(S)  \right) + \int d^2 \bar \theta \left( \frac{p}{4} D^2 L \bar S + \bar W (\bar S )  \right)  .  \label{FO_rewritten}
\end{align}
Substituting (\ref{SL}) and (\ref{SbarL}) into (\ref{FO_rewritten}) one obtains the dual theory in terms of the superfield $L$
\begin{align}
\mathcal L_{L} =&  \int d^4 \theta  \mathcal F(L)   \notag \\
&+ \int d^2 \theta \left( \left( \frac{p}{4b} \bar D^2 L\right)  \ln \left( \frac{-p}{4ab} \bar D^2 L \right) - \frac{p}{4b} \bar D^2 L   \right) \notag \\
&+ \int d^2 \bar \theta \left( \left( \frac{p}{4b} D^2 L\right)  \ln\left( \frac{-p}{4ab} D^2 L \right) - \frac{p}{4b}  D^2  L   \right) . \label{L_L1}
\end{align}
Note that $L$ is a real vector superfield instead of linear superfield. However, as we show in appendix \ref{dofL}, the vector multiplet $L$ contains auxiliary fields that can be eliminated by their equations of motion, and the physical content of the theory can still be described in terms of only a real scalar $l$, a Majorana fermion $\chi$ and a field strength vector field $v_\mu$. 
In fact, a more natural description of this theory will be given by a real scalar $l$, a Majorana fermion $\chi$ and a  pseudoscalar $w$, which is the phase of the $\theta^2$-component of $L$. 
In appendix \ref{dofL}, we compute the Lagrangian in components. The result is
\begin{align}
\mathcal L =&- \frac{1}{4} \mathcal F'(l) \Box l +  \frac{i}{4} \mathcal F''(l) \left( \chi \sigma^\mu \partial \bar \chi + \bar \chi \bar \sigma^\mu \partial_\mu \chi \right) + \frac{a^2 b^2}{p^2} \mathcal F''(l) e^{\frac{b}{p} \mathcal F'(l)} \notag \\
&+ \frac{1}{\mathcal F''(l)} \left( \frac{- \mathcal F'''(l)}{2} \chi \sigma^\mu \bar \chi \partial_\mu w +\frac{p^2}{b^2} \partial^\mu w\partial_\mu w \right) + \left( \frac{\mathcal F''''(l)}{16} - \frac{\mathcal F'''(l)^2}{32 \mathcal F''(l)}\right) \chi^2 \bar \chi^2 \notag \\
&+ \frac{ab}{4p} e^{\frac{b}{p} \mathcal F'(l)/2} \left( \chi^2 e^{-iw} + \bar \chi^2 e^{iw}\right) \left( \mathcal F'''(l) - b \mathcal F''(l)^2 \right) . \label{L_result}
\end{align}
Note that the resulting scalar potential for the field $l$ (last term of the first line of the above expression) is:
\begin{align}
\mathcal V &=  -\frac{a^2 b^2}{p^2} \mathcal F''(l) e^{\frac{b}{p} \mathcal F'(l)} = \frac{a^2 b^2}{p} l^{-2} e^{\frac{b}{l} } .
\end{align}
As a check, we can substitute ({\ref{SdualL2}}) into the above equation to find
\begin{align}
\mathcal V&=\frac{a^2 b^2}{p} (s + \bar s)^2 e^{b (s + \bar s)}, \label{Vchiral}
\end{align}
which corresponds to the scalar potential which can be directly obtained in the chiral formulation from equation (\ref{L_chiral}).

\section{Tensor-scalar duality in supergravity} \label{sugraduality}

We now calculate the same duality in supergravity by using the chiral compensator formalism \cite{Freedman:2012zz}. For this, we start with the Lagrangian
\begin{align}
\mathcal L =& \left[ - \frac{1}{2} \left( S_0 e^{-gV_R} \bar S_0 \right)^3 L^{-2} \right]_D + \left[S_0^3 W(S)\right]_F + \left[f \mathcal W^2 \right]_F + \text{h.c}. \notag \\
&-   \left[L (S + \bar S + c V_R) \right]_D, \label{SUGRALFO}
\end{align}
where, as before, $S$($\bar S$) is an (anti-)chiral multiplet with zero Weyl weight, $L$ is a (unconstrained) real multiplet with Weyl weight 2 and $V_R$ is a vector multiplet of zero Weyl weight associated with a gauged R-symmetry with coupling constant $g$ given by $g = \frac{bc}{3}$. $S_0$ ($\bar S_0$) is the (anti-)chiral compensator superfield with Weyl weight $+1$ ($-1$). The operations $[\ \ ]_D$ and $[\ \ ]_F$ are defined in Appendix \ref{appendixsugra}. 
The superpotential is given by $W(S)=a e^{bS}$ and $\mathcal W^2 = \mathcal W^\alpha \mathcal W_\alpha$ contains the gauge kinetic terms for $V_R$ with gauge kinetic function $f$, which is assumed to be constant $f = \beta$ (at the end of this section we shortly discuss a linear gauge kinetic function).
 
Note that since the R-symmetry generator $T_R$ does not commute with the supersymmetry generators $Q_\alpha$, specifically
\begin{align}
\left[ T_R , Q_\alpha \right] = -i (\gamma_*)_\alpha^{\ \ \beta} Q_\beta,
\end{align}
the global R-symmetry in the previous section necessarily becomes gauged when one couples this theory to gravity. The above Lagrangian thus has a gauged R-symmetry (coupled to the gauged shift symmetry), under which the various fields transform as
\begin{align}
S &\longrightarrow S - ic \Lambda , &  \bar S &\longrightarrow \bar S + ic \bar \Lambda, \notag \\
S_0 & \longrightarrow e^{\frac{ibc}{3} \Lambda} S_0,   &    \bar  S_0  &\longrightarrow  S_0 e^{\frac{-ibc}{3} \bar \Lambda}, \notag \\
V_R &\longrightarrow V_R + i \left( \Lambda - \bar \Lambda \right), \notag \\
L & \longrightarrow L,
\end{align}
 where the gauge parameter $\Lambda$($\bar \Lambda$) is a (anti-)chiral multiplet.

The equations of motion for $L$ are given by
\begin{align}
L = \frac{S_0 e^{-gV_R} \bar S_0  }{(S + \bar S)^{1/3} } . \label{SUGRALS}
\end{align}
In the appendix \ref{appendixsugra}, we fix the conformal gauge by choosing the lowest components $s_0$ and $\bar s_0$ of $S_0$ and $\bar S_0$ as $s_0 \bar s_0 = l^{\frac{2}{3}}$ (see equation (\ref{conformalgauge})), so that the lowest components of equation (\ref{SUGRALS}) read
\begin{align}
l = \frac{1}{s + \bar s}. \label{SUGRAls}
\end{align}
If one substitutes equation (\ref{SUGRALS}) back into the Lagrangian (\ref{SUGRALFO}), one retrieves
\begin{align}
\mathcal L = \left[ - \frac{3}{2} S_0  e^{-gV_R} \bar S_0  (S + \bar S + cV_R )^{2/3}\right]_D + \left[S_0^3 W(S)\right]_F + \left[f \mathcal W^2 \right]_F +  \text{h.c.},
\end{align}
which corresponds to a theory with K\"ahler potential $\mathcal K = -2 \ln \left( S + \bar S \right) $, a superpotential $W(S)$ and a gauged (shift) R-symmetry in the old minimal formalism.

We now couple the Lagrangian (\ref{L_L1}) to supergravity and confirm that the propagating degrees of freedom are still two real scalars and a Majorana fermion, as was the case in rigid supersymmetry.
This is done using the chiral compensator formalism.

To obtain the theory in terms of the superfield $L$, we use the following identity which is proven in Appendix \ref{Proof}
\begin{align}
\left[ L(S + \bar S)\right]_D = \left[ T(L) S \right]_F + \text{h.c.}, \label{D-F}
\end{align}
where $T$ is the chiral projection operator defined below. 

In global supersymmetry, the chiral projection is in superspace the operator $\bar D^2$, where $\bar D_{\dot \alpha}$ is defined in equation (\ref{chiralprojection}). 
In supergravity one can only define such an operator $T$ on a real multiplet with Weyl weight $w$ and chiral weight $c$ given by $(w,c)=(2,0)$. 
Consider now a real multiplet $L$ with components $L = (l, \chi, Z, v_a, \lambda, D)$, where $a$ from now on is used to indicate a flat frame index, 
related to a curved Lorentz index $\mu$ by $v_a = e_a^{\ \mu} v_\mu$. 
Since a chiral multiplet $\Sigma$ can be defined by its lowest component, we define the chiral projection of $L$ via the operator $T$ as a chiral multiplet with lowest component $-\bar Z$
\begin{align}
T(L) \ : \ L \ (2,0) \ \longrightarrow \ \Sigma(L) = T(L), \ \ \ (w,c) = (3,3). \label{T(L)}
\end{align}
In global supersymmetry, this definition of the chiral projection operator $T(L)$ would correspond to $T(L) =- \frac{1}{4} \bar D^2 L$.

By using the relation (\ref{D-F}), one can now calculate the equation of motion for $S$
\begin{align}
S = \frac{1}{b} \ln \left( \frac{T(L)}{ab S^3_0} \right),
\end{align}
and insert this back into the original Lagrangian (\ref{SUGRALFO}) to obtain
\begin{align}
\mathcal L =  & \left[ - \frac{1}{2} \left( S_0 e^{-gV_R} \bar S_0 \right)^3 L^{-2} \right]_D - \left[ c L V_R \right]_D \notag \\& + \frac{1}{b} \left[ T(L)- T(L) \ln \left( \frac{ T(L)}{ab S^3_0} \right) \right]_F + \left[ f \mathcal W^2 \right]_F+ \text{h.c.} \label{SUGRAL}
\end{align}
In Appendix \ref{appendixsugra}, it is shown that apart from the obvious extra fields compared to global supersymmetry, namely the graviton, the gravitino, a gauge boson and a gaugino\footnote{Actually, a linear combination of the two 2-component fermions present in this model is the Goldstino which will be eaten by the gravitino according to the super-BEH mechanism, see section \ref{quantified}. }, this theory does not contain any additional degrees of freedom and the (non-gravitational) spectrum can be described in terms of two real scalars $l$ and $w$, where $w$ is defined as $Z = \rho e^{iw}$, and a Majorana fermion $\chi$. Since the gauge fields do not enter in the tensor-scalar Poincar\'e duality, only the F-term scalar potential is calculated in Appendix \ref{appendixsugra}
\begin{align}
\mathcal V_F = a^2 e^{\frac{b}{l}} \left( \frac{b^2}{2} -2bl - l^2 \right) ,
\end{align}
which upon substituting the relation (\ref{SUGRAls}), one indeed finds the correct F-term scalar potential computed with the chiral formulation (see equation (\ref{scalarpot2}) in section \ref{string theory}), namely
\begin{align}
\mathcal V_F =   a^2 e^{b(s+\bar s)} \left(  \frac{b^2}{2}  - \frac{2b}{s+\bar s}  - \frac{1}{(s+ \bar s)^2 }    \right) .
\end{align}

To make a connection with the scalar potential (\ref{scalarpot2}) of the next section, one has to extend the above formalism to include a linear gauge kinetic function $f(s)$. By following the same method as above, but by including an extra term
\begin{align}
\left[ \alpha S \mathcal W^2 \right]_F + \text{h.c.},
\end{align}
that contains the gauge kinetic terms of $V_R$ with gauge kinetic function $f(S) = \alpha S + \beta$ to the Lagrangian (\ref{SUGRALFO}), one finds that the result (\ref{SUGRAL}) is still valid upon the substitution
 \begin{align}
 T(L) \longrightarrow T(L) - \alpha \mathcal W^2.
 \end{align}
The theory dual to the one described in section \ref{string theory} (for $p=2$) is given by
\begin{align}
\mathcal L =  & \left[ - \frac{1}{2} \left( S_0 e^{-gV_R} \bar S_0 \right)^3 L^{-2} \right]_D - \left[ c L V_R \right]_D \notag \\
& + \frac{1}{b} \left[ \left( T(L) - \alpha \mathcal W^2 \right) - \left( T(L) - \alpha \mathcal W^2 \right) \ln \left( \frac{ T(L) - \alpha \mathcal W^2 }{ab S^3_0} \right) \right]_F + \left[ \beta \mathcal W^2 \right]_F+ \text{h.c.} \label{SUGRAL2}
\end{align}

\section{Scalar potential and connection with string theory} \label{string theory}

From a phenomenological point of view, the scalar potential (\ref{Vchiral}) is runaway and thus not very interesting. However, a supergravity theory with K\"ahler potential (\ref{Kahlerpot}) and superpotential (\ref{superpotential}) can give rise to a a minimum with a parametrically small positive vacuum energy for appropriate choices of the parameters. This was already noticed in \cite{Villadoro:2005yq} and is quantified in the next section for $p=2$.
The important difference with global supersymmetry is that in the local case one can gauge the R-symmetry in order to allow for Fayet-Iliopoulos terms. Moreover, the presence of R-symmetry severely restricts the form of the scalar potential.

The scalar potential is given by \cite{Cremmer:1982en} (we use the conventions of \cite{Freedman:2012zz} and put $\kappa = 1$)
\begin{align}
\mathcal V &= e^{ \mathcal K} \left( -3W \bar W + \nabla_\alpha W g^{\alpha \bar \beta} \bar \nabla_{\bar \beta} \bar W  \right) +  \frac{1}{2} \left( Re(f(s)) \right)^{-1  AB} \mathcal P_A \mathcal P_B  
\end{align}
with covariant derivatives defined by
\begin{align}
\nabla_\alpha W(z) = \partial_\alpha W(z) + \left( \partial_\alpha \mathcal K \right) W(z),
\end{align}
where z denotes the chiral superfields.
$f(s)$ is the gauge kinetic function, which we assume to be $f(s) = s + d$ (as in the Heterotic string effective action), where we take $d$ to be real;  this is the most general gauge kinetic function allowed by the shift symmetry.  $\mathcal P_A$ are the moment maps
\begin{align}
\mathcal P_A =  i\left( k_A^\alpha \partial_\alpha \mathcal K - r_A \right), \label{momentmaps}
\end{align}
with $k_A$ the Killing vectors, i.e. $\delta z^\alpha = \theta^A k_A^\alpha(z)$. The functions $r_A(z)$ satisfy
\begin{align}
W_\alpha k_A^\alpha = - r_A W
\end{align}
and give rise to a Fayet-Iliopoulos term (see equation (\ref{momentmaps})) if $r_A$ is a complex constant. 
For the supergravity theory with K\"ahler potential (\ref{Kahlerpot}) and superpotential (\ref{superpotential}), invariant under the shift symmetry $\delta s =  \theta k = - ic \theta $,  the scalar potential is given by
\begin{align}
\mathcal V =& \frac{-3 a^2 }{(s + \bar s)^p} e^{b(s + \bar s)} + \left( b - \frac{p}{s + \bar s} \right)^2 \left( \frac{a^2}{p} \frac{e^{b(s + \bar s)}}{(s + \bar s)^{p-2}} + \frac{c^2}{s + \bar s + 2d} \right) . \label{scalarpot}
\end{align}
For example, for $p=2$, this gives
\begin{equation}
\mathcal V =  a^2 e^{b(s+\bar s)} \left(  \frac{b^2}{2}  - \frac{2b}{s+\bar s}  - \frac{1}{(s+ \bar s)^2 }    \right) 
+  \frac{c^2}{s + \bar s + 2d} \left( \frac{2}{s+\bar s} - b \right)^2.  \label{scalarpot2}
\end{equation}
Taking $b=d=0$, this reduces to 
\begin{equation}
\mathcal V =  -  \frac{a^2}{(s+ \bar s)^2 }   +  \frac{4 c^2}{(s+\bar s)^3}, \label{Vb0}
\end{equation}
The potential (\ref{Vb0}) coincides with the one derived in \cite{Antoniadis:2008uk} from D-branes in non-critical strings.
Indeed the second term corresponds to a disk contribution proportional to the D-brane tension deficit $\delta \bar T$ induced by the presence of magnetized branes in type I orientifold compactifications, while the first term corresponds to an additional contribution at the sphere-level, induced from going off criticality, proportional to the central charge deficit $\delta c$. 
More precisely, in the Einstein frame ($M_p = 1$), the scalar potential in \cite{Antoniadis:2008uk} is given by
\begin{align}
\mathcal V_{nc} = e^{2\varphi_4} \delta c + \frac{e^{3 \varphi_4}}{(2 \pi)^3 v_6^{1/2}} \delta \bar T, \label{V_nc}
\end{align}
where $\varphi_4$ is the four-dimensional dilaton related to the ten-dimensional dilaton $\varphi$ by $e^{-2 \varphi_4} = e^{-2 \varphi} v_6$, and $v_6$ is the six-dimensional volume given by $v_6 = \frac{V_6}{(4 \pi^2 \alpha ')^3}$. 
By identifying $e^{-\varphi_4} = \text{Re}(s)$, one sees that the scalar potential (\ref{V_nc}) is indeed of the form of equation (\ref{Vb0}). One can then identify $\delta c$ and $\delta \bar T$ as 
\begin{align}
\delta c &= -\frac{a^2}{4} \\
\delta \bar T &= \frac{(2 \pi)^3 v_6^{1/2}}{2} c^2 .
\end{align}
Note that $\delta c$ can become infinitesimally small only if it is negative, as is needed for the existence of an anti-de Sitter vacuum in equation (\ref{V_nc}). The potential has a minimum at $s_0 + \bar s_0 = \frac{6c^2}{a^2}$, given by $\mathcal V(s_0) = -\frac{a^6}{108 c^4} < 0$.

It is interesting to note that a nonzero $b$ implied by the most general superpotential consistent with the shift symmetry allows one to find (meta-stable) de-Sitter vacua for certain values of the parameters, as we show below.

\section{Meta-stable de Sitter vacua quantified.} \label{quantified}

The scalar potential (\ref{scalarpot}) with $d=0$ and $p=2$ was already examined in  \cite{Villadoro:2005yq}, where it was noticed that for $b >0$ there is always a stable supersymmetric anti-deSitter (AdS) vacuum. 
The authors also noticed that for $b < 0$, there is a locally stable de Sitter (dS) minimum of the potential for a range of values of the parameters.
In this section we quantify these results, compute the spectrum and show that the masses of all physical fields are of the order of the supersymmetry breaking scale.

The fact that for $b>0$, there always exists a stable AdS vacuum, can be seen by considering the derivative of the scalar potential, which is given by
\begin{align}
\mathcal V' =&  \left( b - \frac{p}{s + \bar s} \right) \left( \frac{-3 a^2}{(s + \bar s)^p} e^{b(s + \bar s)} + 2p \left( \frac{a^2}{p} \frac{e^{b(s + \bar s)}}{(s + \bar s)^p} + \frac{c^2}{(s + \bar s + 2d)(s + \bar s)^2} \right) \right) \notag \\
&+ \left( b - \frac{p}{s + \bar s} \right)^2 \left( \frac{(2-p) a^2}{p} \frac{e^{b(s + \bar s)}}{(s + \bar s)^{p-1}} - \frac{ba^2}{p} \frac{e^{b(s + \bar s)}}{(s + \bar s)^{p-2}} - \frac{c^2}{(s + \bar s + 2d)} \right).
\end{align}

It is clear that for $b>0$, the potential has a minimum at $b = \frac{p}{s_0 + \bar s_0}$. The F-term and the D-term contributions to the scalar potential ($\mathcal V = \mathcal V_{SG} + \mathcal V_F + \mathcal V_D$, where $\mathcal V_{SG} = -3 e^{\mathcal K} W\bar W$)
\begin{align}
\mathcal V_F &= e^{ \mathcal K} \nabla_\alpha W g^{\alpha \bar \beta} \bar \nabla_{\bar \beta} \bar W =  \frac{a^2}{p}  \frac{e^{b(s + \bar s)}}{(s + \bar s)^{p-2}}  \left( b - \frac{p}{s + \bar s} \right)^2 = 0,\notag \\
\mathcal V_D &=  \frac{1}{2} \left( Re(f(s)) \right)^{-1  AB} \mathcal P_A \mathcal P_B = \frac{ c^2}{s + \bar s + 2d}  \left( b - \frac{p}{s + \bar s} \right)^2 = 0,
\end{align}
both vanish in this vacuum, thus confirming that supersymmetry is unbroken.

In the context of string theory, the case of $b>0$ is unphysical while $b<0$ corresponds to a non-perturbative superpotential, completely fixed from the requirement of the shift symmetry or the gauged R-symmetry.
We dedicate the rest of this section to showing explicitly that one can indeed find a parametrically small and positive vacuum energy for the case $b<0$ and $p=2$.
For this we first look for a Minkowski minimum and solve the equations
\begin{align}
\mathcal V(s_0) &= 0, \\
\frac{d \mathcal V}{ds}(s_0) &= 0,
\end{align}
where $s_0$ is the value of $s$ at the minimum of the potential. This leads to the following relations between the parameters
\begin{align}
b (s_0 + \bar s_0)= \alpha \approx -0.183268 \label{bsbars},\\
\frac{a^2}{b c^2} = A(\alpha) \approx -50.6602 \label{V_minkowski},
\end{align}
 where $\alpha$ is the root of the polynomial $- x^5 + 7 x^4 - 10 x^3 - 22 x^2 + 40 x + 8$ close to  $-0.18$ and $A(\alpha)$ is given by 
\begin{align}
 A(\alpha) = \frac{e^{-\alpha}}{ \alpha} \left( \frac{-4 + 4 \alpha - \alpha^2}{\frac{\alpha^2}{2} - 2 \alpha - 1} \right). \label{Aalpha}
\end{align}
 Note that the above polynomial has five roots, four of which are unphysical: two roots are imaginary, one is positive, which is incompatible with equation (\ref{bsbars}), since $s_0 + \bar s_0$ should be positive (see equation (\ref{sgaugecoupling})), and a fourth root gives rise to a positive $A(\alpha)$, which is incompatible with equation (\ref{V_minkowski}).
Note that the position of the minimum (and thus the value of the string coupling constant $g_s$) is only determined by $b$.

The gravitino mass term is given by
\begin{align}
(m_{3/2})^2 &= e^{\mathcal G} \notag \\ &= \frac{a^2}{s_0 + \bar s_0} e^{b(s_0 + \bar s_0)} \notag \\ &=\frac{a^2 b^2}{\alpha^2} e^{\alpha} , \label{m32}
\end{align}
where $\mathcal G = \mathcal K + \ln(W \bar W)$ and we used equation (\ref{bsbars}) to go from the second to the third line. This shows that for nonzero $a$ and $b<0$ supersymmetry is broken.
Indeed the F-term and D-term contributions to the scalar potential are now 
\begin{align}
\left. \mathcal V_F \right|_{s + \bar s = \frac{\alpha}{b}} &= \frac{1}{2} a^2 b^2 e^\alpha \left( 1 - \frac{2}{\alpha} \right)^2 > 0, \notag \\
\left. \mathcal V_D \right|_{s + \bar s = \frac{\alpha}{b}} &= \frac{b^3 c^2}{\alpha}  \left( 1 - \frac{2}{\alpha} \right)^2 > 0 .
\end{align}
Note that for the D-term, both $\alpha$ and $b$ are negative, so that the D-term contribution to the superpotential is indeed positive.

Due to the Stueckelberg coupling, the imaginary part of $s$ (the axion) gets eaten by the gauge field, which acquires a mass. On the other hand, the Goldstino, which is a linear combination of the fermion of the chiral multiplet $\chi$ and the gaugino $\lambda$ gets eaten by the gravitino.
As a result, the physical spectrum of the theory consists (besides the graviton) of a massive scalar, namely the dilaton, a Majorana fermion, a massive gauge field and a massive gravitino.
A calculation of the masses of these fields is given in Appendix \ref{masses}. The results are
 \begin{align}
 m_s &=  \frac{b^2 c}{\alpha^2} \sqrt{ \frac{  48+192 \alpha-128 \alpha^2-8 \alpha^3+24 \alpha^4-8 \alpha^5+\alpha^6 } {2+4 \alpha-\alpha^2} }, \\
  m_{A_\mu} &= \frac{2c}{s + \bar s}, \\
m_f^2 &=  a^2 b^2 \frac{e^\alpha \left( 116 + 68 \alpha - 15 \alpha^2  - 4 \alpha^3 + 8 \alpha^4\right)}{144 \alpha^2}. 
 \end{align}
 Note that all masses are of the same order of magnitude, since they are proportional to the same constant $a$ (or $c$ related by eq.(\ref{V_minkowski}) where $b$ is fixed by eq.(\ref{bsbars})), which is a free parameter of the model. Thus, they vanish in the same way in the supersymmetric limit $a\to 0$.

\begin{figure}[h!]
    \centering
    \includegraphics[width=0.9\textwidth]{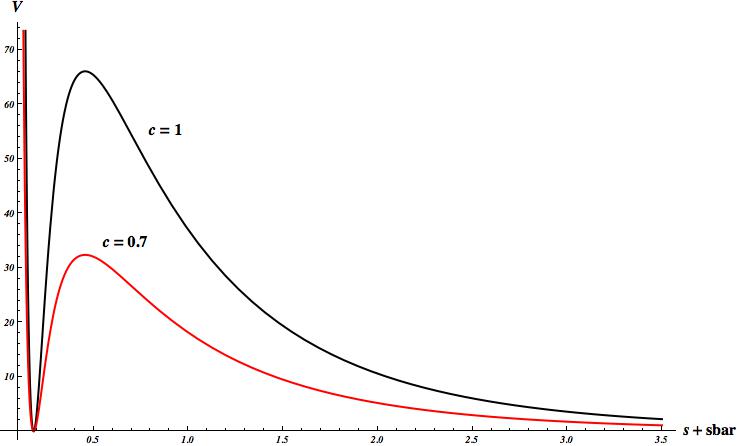}
    \caption{A plot of the scalar potential for $p=2$, $b=-1$, $d=0$, and $a$ given by equation (\ref{V_minkowski}) for $c=1$ (black curve) and $c=0.7$ (red curve).} 
    \label{fig:potential}
\end{figure}

If one were to shift the value of the minimum of the potential to a small positive value $\Lambda$, (\ref{V_minkowski}) gets modified to\begin{align}
 A_\Lambda (\alpha)  = \frac{a^2}{bc^2} - \frac{\Lambda}{b^3 c^2} \left( \frac{\alpha^2 e^{-\alpha}}{\frac{\alpha^2}{2} - 2\alpha -1}  \right) .
\end{align}
It follows that by carefully tuning $a$ and $c$, $\Lambda$ can be made positive and arbitrarily small independently of the supersymmetry breaking scale. A plot of the scalar potential for certain values of the parameters is shown in figure \ref{fig:potential}.

\section{Conclusions}
In this work, we studied a simple supergravity model that has a de Sitter metastable vacuum with an arbitrarily tunable value of the cosmological constant and supersymmetry breaking by an expectation value of an F and D-term. The two scales, vacuum energy and supersymmetry breaking, are independent parameters of the theory. The minimal spectrum of the model consists of a vector multiplet that gauges an R-symmetry and a single chiral multiplet whose pseudoscalar component transforms by a shift under the R-symmetry. The form of both the superpotential and the FI D-term is fixed by the symmetry, up to two arbitrary constants corresponding to the two free parameters of the theory. The chiral multiplet can be described equivalently by a linear multiplet using Poincar{\' e} duality, even in the presence of the superpotential.

The model can in principle be embedded in (type I) string theory, upon identification of the chiral (or linear) multiplet with the dilaton or a compactification modulus. The question of an exact gauged (shift) R-symmetry requires however further investigation. This model can obviously be extended in the presence of more fields and it would be interesting to study possible phenomenological applications in string model building with moduli stabilization at a supersymmetry breaking vacuum with arbitrarily small positive cosmological constant.

\section*{Acknowledgements}
We thank S. Ferrara, J.-C. Jacot, R. Monten and A. Van Proeyen for useful discussions. This work was supported in part by the European Commission under the ERC Advanced Grant 226371. 

\appendix
\section{Propagating degrees of freedom of $L$ in (\ref{L_L1})} \label{dofL}

Define the general expression of the real superfield $L$ as follows
\begin{align}
L =& l +i \theta \chi - i \bar \theta \bar \chi + \theta^2 Z + \bar \theta^2 \bar Z - \theta \sigma^\mu \bar \theta v_\mu \notag \\ &+ i \theta^2 \bar \theta (\bar \lambda - \frac{i}{2}\bar \sigma^\mu \partial_\mu \chi) -  i \bar \theta^2 \theta ( \lambda - \frac{i}{2} \sigma^\mu \partial_\mu \bar \chi) + \frac{1}{2} \theta^2 \bar \theta^2 (D - \frac{1}{2} \Box l) ,
\end{align}
Rewriting $Z = \rho e^{iw}$ and $\bar Z = \rho e^{-iw}$, it follows that the Lagrangian (\ref{L_L1}) is given in components by
\begin{align}
\mathcal L =& \frac{\mathcal F'(l)}{2} \left( D - \frac{1}{2} \Box l \right) \notag \\
 & + \frac{\mathcal F''(l)}{2} \left( -\chi  \left( \lambda - \frac{i}{2} \sigma^\mu \partial_\mu \bar \chi \right) - \bar \chi \left( \bar \lambda - \frac{i}{2} \bar \sigma^\mu \partial_\mu \chi \right)  + 2 \rho^2 - \frac{1}{2} v^\mu v_\mu \right) \notag \\
 & +\frac{1}{4} \mathcal F'''(l) \left( \chi^2 \rho e^{-iw} + \bar \chi^2 \rho e^{iw} + \chi \sigma^\mu \bar \chi v_\mu \right) + \frac{\mathcal F''''(l)}{16} \chi^2 \bar \chi^2  \notag \\
&  + \frac{p}{b} D \ln \left( \frac{p}{ab} \rho \right) + \frac{p}{b} w \partial^\mu v_\mu  -  \frac{p}{4b} \frac{\lambda^2}{\rho}e^{iw} -  \frac{p}{4b} \frac{\bar \lambda^2}{\rho}e^{-iw} . \label{LF_L}
\end{align}
The equations of motion for the fields $D,\lambda, v_\mu$ and $\rho$ are
\begin{align}
\rho &= \frac{ab}{p} e^{-\frac{b}{2p}(\mathcal F '(l))}, \notag \\
p \frac{\lambda}{b \rho} e^{iw} &= - \mathcal F''(l) \chi, \notag \\
\frac{\mathcal F''}{2} v_\mu &=  \frac{\mathcal F'''}{4} \chi \sigma^\mu \bar \chi - \frac{p}{b}\partial^\mu w , \notag \\
-\frac{p}{b} D &=2 \mathcal F'' \rho^2 + \frac{\rho}{4} \mathcal F''' \left( \chi^2 e^{-iw} + \bar \chi^2 e^{iw}\right) + \frac{p}{4b} (\frac{\lambda^2}{\rho}e^{iw} + \frac{\bar \lambda^2}{\rho}e^{-iw}), \end{align}
and can be used to eliminate these fields from the Lagrangian by substituting them back into (\ref{LF_L}) 
\begin{align}
\mathcal L =&- \frac{1}{4} \mathcal F'(l) \Box l +  \frac{i}{4} \mathcal F''(l) \left( \chi \sigma^\mu \partial \bar \chi + \bar \chi \bar \sigma^\mu \partial_\mu \chi \right) + \frac{a^2 b^2}{p^2} \mathcal F''(l) e^{\frac{b}{p} \mathcal F'(l)} \notag \\
&+ \frac{1}{\mathcal F''(l)} \left( \frac{- \mathcal F'''(l)}{2} \chi \sigma^\mu \bar \chi \partial_\mu w +\frac{p^2}{b^2} \partial^\mu w\partial_\mu w \right) + \left( \frac{\mathcal F''''(l)}{16} - \frac{\mathcal F'''(l)^2}{32 \mathcal F''(l)}\right) \chi^2 \bar \chi^2 \notag \\
&+ \frac{ab}{4p} e^{\frac{b}{p} \mathcal F'(l)/2} \left( \chi^2 e^{-iw} + \bar \chi^2 e^{iw}\right) \left( \mathcal F'''(l) - \frac{b}{p} \mathcal F''(l)^2 \right). \label{L_result}
\end{align}
It follows that the Lagrangian can be written only in terms of the physical propagating fields $l, \chi$ and $w$.
Note that it is also possible to write the Lagrangian in terms of $l, \chi$ and $v_\mu$. The above calculation proves that the field strength $v_\mu$ contains only one degree of freedom.

\section{Supergravity} \label{appendixsugra}

We now look for a component expression of the Lagrangian (\ref{SUGRAL}), repeated here for convenience
\begin{align}
\mathcal L =&   \left[ - \frac{1}{2} \left( S_0 e^{-gV_R} \bar S_0 \right)^3 L^{-2} \right]_D - \left[ c L V_R \right]_D \notag \\&+ \frac{1}{b} \left[ T(L)- T(L) \ln \left( \frac{ T(L)}{ab S^3_0} \right) \right]_F + \left[f \mathcal W^2 \right]_F +\text{h.c.}
\end{align}
Since we are interested in comparing the F-term scalar potential of the above theory, we will from now on neglect the vector multiplet $V_R$ putting $g=c=f=0$.
The components of the real multiplet $L$ are given by $L = (l, \chi, Z, v_a, \lambda, D)$, while the components of the (auxiliary) chiral multiplet are given by $S_0 = (s_0, P_L \zeta, F)$, where $P_{L(R)}$ is the left-handed (right-handed) projection operator.
The Weyl and chiral weights of the $Z$-component of a linear multiplet are given by $(w,c) = (w+1, -3)$. Since the lowest component of a chiral multiplet must satisfy $w=c$ for the superconformal algebra to close, it is clear that the operation $T(L)$, given by equation (\ref{T(L)}), can only be defined on a real multiplet with $w=2$, so that $\bar Z$ has weights $(w,c)=(3,3)$ and can thus be used as lowest component of a chiral multiplet.  This agrees with the assumption that $L$ has Weyl weight 2. The auxiliary field $S_0$ is chosen to have weight 1. In the following we use the conventions of \cite{Freedman:2012zz}.

The chiral multiplet obtained by the projection operator $T$ has components $T(L) = (- \bar Z, -\sqrt 2 i P_L \left(\lambda + \cancel{ \mathcal D} \chi \right), D + \mathcal D^a \mathcal D_a l + i \mathcal D^a v_a)$, where the covariant derivatives are given by
\begin{align}
\mathcal D_a l &= \partial_a l - 2 \mathcal{B}_a l - \frac{1}{2} i \bar \psi_a \gamma_* \chi, \notag \\
\mathcal D_a v_b &= \partial_a v_b - 3 \mathcal B_a v_b + \frac{1}{2} \bar \psi_a \left( \gamma_b \lambda + \mathcal D_b \chi \right) - \frac{3}{2} \bar \varphi_a \gamma_b \chi, \notag \\
\mathcal D_a P_L \chi &= \left( \partial_a + \frac{1}{4} \omega_a^{\ bc} \gamma_{bc} - \frac{5}{2} \mathcal B_a + \frac{3}{2} i \mathcal A_a \right) \chi - \frac{1}{2} P_L \left( i Z - \cancel v - i \cancel {\mathcal D} l \right) \psi_a + 2i l P_L \varphi_a,
\end{align}
where $\psi_\mu$ is the gravitino, $\varphi_\mu$ is the gauge field corresponding to the conformal supercharge $\mathcal S$, $\mathcal B_\mu$ is the gauge field of dilatations and $\mathcal A_\mu$ is the gauge field of the T-symmetry (which is the $U(1)$ R-symmetry of the superconformal algebra).
An F-term Lagrangian is defined on a chiral multiplet $X=(s, P_L \zeta, F)$ with weight 3
\begin{align}
\left[ X \right]_F = e  \ \text{Re} \left( F + \frac{1}{\sqrt 2} \bar \psi_\mu \gamma^\mu P_L \zeta + \frac{1}{2} s \bar \psi_\nu \gamma^{\mu \nu} P_R \psi_\nu \right), \label{F-term}
\end{align}
where $e$ is the determinant of the vierbein. From now on, we write $P_{L,R} \zeta = \zeta_{L,R}$.
A D-term Lagrangian is defined on a real multiplet $C = (C, \chi, Z, v_a, \lambda, D)$ with weight 2 as
\begin{align}
 \left[ C \right]_D = &e \ \left( D - \frac{1}{2} \bar \psi \cdot \gamma i \gamma_* \lambda - \frac{1}{3} C R(\omega) + \frac{1}{6} \left( C \bar \gamma_\mu \gamma^{\mu \rho \sigma } - i \bar \chi \gamma^{\rho \sigma} \gamma_* \right) R'_{\rho \sigma}(Q) \right. \notag \\ &\left. + \frac{1}{4} \epsilon^{abcd} \bar \psi_a \gamma_b \psi_c \left( v_d - \frac{1}{2} \bar \psi_d \chi \right) \right), \label{D-term}
\end{align}
where $R(\omega)$ and $R'(Q)$ are the graviton and gravitino curvatures.
To obtain a real multiplet from a real function $\phi (s, \bar s)$ of chiral superfields $(s, \zeta_L, F)$ and their anti-chiral counterparts, we use \cite{VanProeyen:1983wk}
\begin{align}
&\left(  \phi (s, \bar s) ; - \sqrt 2 i \phi_s \zeta_{L}  ; - 2 \phi_s F + \phi_{ss} \bar \zeta_L \zeta_L ; i \phi_s \partial_a s - i \phi_{\bar s} \partial_a \bar s + i \phi_{s \bar s} \bar \zeta_L \gamma_a \zeta_R ; \right.  \notag \\
& -\sqrt 2 i \phi_{s \bar s} F \zeta_R + \sqrt 2 i \phi_{s \bar s} \cancel \partial \bar s \zeta_L + \frac{i}{\sqrt 2} \phi_{s s \bar s} \zeta_R \cdot \bar \zeta_L \zeta_L ; \notag \\
&\phi_{s \bar s} \left[ 2 F \bar F - 2 \partial_a s \partial^a \bar s - \partial_a \bar \zeta_L \gamma^a \zeta_R + \bar \zeta_L \cancel \partial \zeta_R \right] - \phi_{s s \bar s} \left[ \bar \zeta_L \zeta_L \bar F + \bar \zeta_L \cancel \partial s \zeta_R \right] \notag \\
&\left.  - \phi_{s \bar s \bar s} \left[ \bar \zeta_R \zeta_R F + \bar \zeta_R \cancel  \partial \bar s \zeta_L \right] + \frac{ \phi_{s s \bar s \bar s}}{2} \bar \zeta_L \zeta_L \cdot \bar \zeta_R \zeta_R \right), \label{chiralcomponents}
\end{align}
where $\phi_s = \frac{\partial \phi}{\partial s}$, $\phi_{\bar s} = \frac{\partial \phi}{\partial \bar s}$, etc.
To obtain a real superfield from a function $f(C)$ of real multiplets with components $C_i, \chi_{Li}, Z_i, v_{ai}, \lambda_{Ri}, D_i$ we use
\begin{align}
&\left( f(C); f^i \chi_{Li} ;  f^i Z_i - \frac{1}{2} f^{ij} \bar \chi_{Li} \chi_{Lj} ; f^i v_{ai} + \frac{i}{2} f^{ij} \bar \chi_{Li} \gamma_a \chi_{Rj}; \right. \notag \\
& f^i \lambda_{Ri} + \frac{1}{2} P_L \left[ Z_i -  i \cancel v_i - \cancel \partial C_i \right] \chi_j - \frac{1}{4} f^{ijk} \chi_{Ri} \cdot \bar \chi_{Lj} \chi_{Lk} ; \notag \\
& f^i D_i - \frac{1}{2} f^{ij} \left[ 2 \bar \chi_i \lambda_j - Z_i \bar Z_j + v_{ai} v^a_j + \partial_a C_i \partial^a C_j - \left( \partial_a \bar \chi_{Li} \right) \gamma^a \chi_{Rj} + \bar \chi_{Li } \cancel \partial \chi_{Rj} \right] \notag \\
& \left. - \frac{1}{4} f^{ijk} \left[ \bar \chi_{Li} \bar Z_j \chi_{Lk} + \bar \chi_{Ri} Z_j \chi_{Rk} + 2i \bar \chi_{i} \cancel v_j \chi_k  \right] + \frac{1}{8} f^{ijkl} \bar \chi_{Li} \chi_{Lj} \cdot \bar \chi_{Rk} \chi_{Rl} \right). \label{realcomponents}
\end{align}

We are now ready to write down the (non-gauge) bosonic part of the Lagrangian (\ref{SUGRAL}) as
\begin{align}
\mathcal L/e &=  \frac{1}{2} (s_0 \bar s_0)^3 l^{-2} R +  (s_0 \bar s_0)^{3} l^{-3} D  + \frac{3}{2}  (s_0 \bar s_0)^{3} l^{-4}  v^\mu v_\mu  \notag \\
& - \frac{1}{b} \ln \left( \frac{\rho^2}{a^2 b^2 (s_0 \bar s_0)^3} \right) D + \frac{1}{b} 2 v^\mu \partial_\mu w   + \mathcal L_{kin} - \mathcal V_F, \label{VkinVF}
\end{align}
where we used that 
\begin{align}
\Box l = \mathcal D^a \mathcal D_a l = \partial^a \partial_a l - \frac{1}{3} l R \label{boxl}
\end{align}
 to rewrite a term involving $\Box l$ in the third term in (\ref{SUGRAL}). The scalar field $w$ is defined as 
\begin{align}
Z = \rho e^{iw}, \label{Zw}
\end{align}
 and it will turn out that this field inherits the gauged shift symmetry.
$\mathcal L_{kin}$ and $\mathcal V_F$ in equation (\ref{VkinVF}) are given by
\begin{align}
\mathcal L_{kin} &=   +3  (s_0 \bar s_0)^{2} l^{-2}   \partial^\mu s_0 \partial_\mu \bar s_0    + \frac{3}{2}  (s_0 \bar s_0) l^{-2} \partial^\mu(s_0 \bar s_0) \partial_\mu (s_0 \bar s_0)   -3   (s_0 \bar s_0)^{2} l^{-3}  \partial^\mu l \partial_\mu (s_0 \bar s_0) \notag \\ 
& + \frac{3}{2}  (s_0 \bar s_0)^{3} l^{-4}  \partial^\mu l \partial_\mu l  -\left( \frac{s_0 s_0}{l} \right)^3 \partial^\mu \partial_\mu l   \notag \\
\mathcal V_F &=  9   (s_0 \bar s_0)^{2} l^{-2}  F \bar F    + 3   (s_0 \bar s_0)^{2} l^{-3} \left( \bar s_0 Z \bar F + s_0 \bar Z F \right)  + 3 \frac{Z \bar F}{b \bar s_0} + 3 \frac{\bar Z F}{b s_0}  +  \frac{3}{2}  (s_0 \bar s_0)^{3} l^{-4} Z \bar Z \notag \\
&= - \frac{\rho^2}{(s_0 \bar s_0)^3} l^2 \left( \frac{1}{b} +  (s_0 \bar s_0)^2  l^{-3} \right)^2  +  \frac{3}{2} \rho^2 (s_0 \bar s_0)^{3} l^{-4} ,
\end{align}
where we solved the equations of motion of the auxiliary field $F$ and used equation (\ref{Zw}) to go from the first to the second line of $\mathcal V_F$. On the other hand, the equations of motion for $D$ and $v_\mu$ are given by
\begin{align}
\rho^2= a^2 b^2 (s_0 \bar s_0)^3 e^{ b \left( \frac{s_0 \bar s_0}{l}\right)^3 }, \notag \\
 (s_0 \bar s_0)^3 l^{-4} v_\mu =  - \frac{2}{3b} \partial_\mu w.
\end{align}
We now fix the conformal gauge as
\begin{align}
(s_0 \bar s_0)^3 = l^2 \label{conformalgauge}
\end{align}
to obtain a correctly normalised Einstein-Hilbert term. Upon this choice $\mathcal L_{kin}$ reduces to $\mathcal L_{kin} = -\frac{1}{2} \frac{1}{l^2} \partial^\mu l \partial_\mu l$ and we obtain the final result
\begin{align}
\mathcal L = \frac{1}{2} R - \frac{1}{2} \frac{1}{l^2} \partial^\mu l \partial_\mu l - \frac{2}{3 b^2} \partial^\mu w \partial_\mu w - \mathcal V_F,
\end{align}
 with 
 \begin{align}
\mathcal V_F = a^2 e^{\frac{b}{l}} \left( \frac{b^2}{2} -2bl - l^2 \right) .
\end{align}

 \section{Proof of an important identity} \label{Proof}
 
 In this Appendix we prove equation (\ref{D-F}), repeated here for convience
 \begin{align}
\left[ L(S + \bar S)\right]_D = \left[ T(L) S \right]_F + \text{h.c.} \label{D-Fappendix}
\end{align}
To do this, we calculate both sides in components and see that they coincide. It is sufficient to show this only for the bosonic terms.
If the (bosonic) components of the chiral multiplet $S$ are given by $\left( s, 0, F \right)$ and the (bosonic) components of the real multiplet $L$ are given 
by $\left( l \ ; 0 \ ; Z \ ; v_\mu \ ; 0 \ ; D \  \right)$, then the components of $S + \bar S$ can be calculated using (\ref{chiralcomponents})
\begin{align}
S + \bar S &= \left(s + \bar s \ ; 0 \ ; -2(F + \bar F) \ ; i \partial_\mu (s - \bar s) \ ; 0 \ ; 0 \  \right).
\end{align}
The chiral multiplet $T(L)$ is defined as a chiral multiplet with lowest component $-\bar Z$ and has components
\begin{align}
T(L) = (- \bar Z \ ; 0 \ ; D + \Box l + i \mathcal D^\mu v_\mu \ ).
\end{align}
The left-hand side of equation (\ref{D-Fappendix}) can be written in components by using equations (\ref{chiralcomponents}) and (\ref{D-term}) (neglecting all terms involving fermions)
\begin{align}
\left[L (S + \bar S) \right]_D = -\frac{1}{3} l (s + \bar s) R + (s + \bar s) D - \bar Z F - Z \bar F - i v^\mu \partial_\mu (s - \bar s) - \partial^\mu \partial_\mu (s + \bar s).
\end{align}
Similarly, by using equation (\ref{chiralcomponents}), $\left[ T(L) S \right]_F$ can be written in components as
\begin{align}
\left[ T(L) S \right]_F = - \bar Z F + s D + s \Box l + i s \partial^\mu v_\mu
\end{align}
The right-hand side is thus given by
\begin{align}
\left[ T(L) S \right]_F + \text{h.c.}  = - \bar Z F - Z \bar F + (s + \bar s) D + (s + \bar s) \Box l + i (s + \bar s) \partial^\mu v_\mu .
\end{align}
The equality (\ref{D-Fappendix}) is then proven by using equation (\ref{boxl}) for $\Box l$ and then partially integrating the last two terms.

 \section{Masses of the scalar, the fermion and the gauge field} \label{masses}

\subsection{Mass of the gauge field}

The kinetic term $\mathcal L_s $ of the scalar field $s$ can be written as
\begin{align}
\mathcal L_s /e = - g_{s \bar s} \hat \partial_\mu s \hat \partial^\mu \bar s, \label{kinetics}
\end{align}
where the covariant derivative is defined as
\begin{align}
\hat \partial_\mu s &= \partial_\mu s - k A_\mu  \notag \\
&= \partial_\mu s + ic A_\mu,
\end{align}
where $k$ is the Killing vector associated with the shift symmetry given by $k = -ic$. Equation (\ref{kinetics}) can then be written as
\begin{align}
\mathcal L_s/e &=  \frac{-2}{(s + \bar s)^2} \left( \partial^\mu s \partial_\mu \bar s - ic A^\mu \partial_\mu(s -\bar s)+ c^2 A^\mu A_\mu \right) .
 \end{align}
 The local shift symmetry allows us to gauge fix $s-\bar s = 0$. The gauge field now acquires a mass term and its mass is given by
 \begin{align}
 m_{A_\mu} = \frac{2c}{s + \bar s}.
 \end{align}
  
\subsection{Mass of the scalar}

We isolate the quadratic contribution of the superpotential, which is repeated here for convenience
\begin{equation}
\mathcal V =  a^2 e^{b(s+\bar s)} \left(  \frac{b^2}{2}  - \frac{2b}{s+\bar s}  - \frac{1}{(s+ \bar s)^2 }    \right) 
+  \frac{c^2}{s + \bar s + 2d} \left( \frac{2}{s+\bar s} - b \right)^2.
\end{equation}
The quadratic term $\mathcal V_{quad} = \frac{1}{2} \mathcal V''(\alpha/b) \left( s + \bar s - \frac{\alpha}{b} \right)^2$ in the series expansion of $\mathcal V$ around its minimum at $s + \bar s = \frac{\alpha}{b}$ is given by
\begin{align}
\mathcal V_{quad}  = \frac{b^4}{4 \alpha^5}  \left(4 \left(24-12 \alpha+\alpha^2\right) b c^2+a^2 \alpha \left(-12+6 \alpha^2-4 \alpha^3+\alpha^4\right) e^{\alpha}\right) \left(s + \bar s - \frac{\alpha}{b}\right)^2 . \notag 
\end{align}
This expression can be simplified by using equation (\ref{V_minkowski})
\begin{align}
\mathcal V_{quad} = -\frac{\left(48+192 \alpha-128 \alpha^2-8 \alpha^3+24 \alpha^4-8 \alpha^5+\alpha^6\right) b^5 c^2 (s + \bar s - \frac{\alpha}{b})^2}{2 \alpha^5 \left(-2-4 \alpha+\alpha^2\right)} .
\end{align}
Taking into account the non-canonical form of the kinetic terms\footnote{
The kinetic term of $\text{Re}(s)$ is given by
\begin{align}
\mathcal L_{\text{kin}}  = \frac{-1}{2\text{Re}(s)^2} \partial_\mu \text{Re}(s) \partial^\mu \text{Re}(s) .
\end{align}
Expanding around the mimimum $\phi = \text{Re}(s) - S_0$, with $S_0 = \frac{\alpha}{2b}$, gives
\begin{align}
\mathcal L = \frac{-1}{2S_0^2} \left( 1 - 2 \frac{\phi}{S_0} +\dots \right) \partial^\mu \phi \partial_\mu \phi .
\end{align}
We should thus rescale the mass with a factor $S_0$.
}, the mass of the dilaton $\text{Re}(s) = \frac{s + \bar s}{2}$ is then given by
\begin{align}
m_s^2 =   \frac{   - b^4 c^2  }{ \alpha^4 \left(-2-4 \alpha+\alpha^2\right)} \left(48+192 \alpha-128 \alpha^2-8 \alpha^3+24 \alpha^4-8 \alpha^5+\alpha^6\right) .
\end{align}
For the earlier calculated numerical value of $\alpha$, this expression is indeed positive: $m_s^2 > 0$.

\subsection{Mass of the fermion and elimination of the Goldstino}

The Goldstino is given by
 \begin{align}
P_L \nu &= \chi^\alpha \delta_s \chi_\alpha + P_L \lambda^A \delta_s P_R \lambda_A,
\end{align}
where $\delta_s \chi_\alpha$ and $\delta_s P_R \lambda_A$ are defined as
\begin{align}
\delta_s \chi_\alpha &= - \frac{1}{\sqrt 2}  e^{K/2} \nabla_\alpha W, \\
\delta_s P_R \lambda_A &= - \frac{1}{2} i  \mathcal P_A.
\end{align}
The indices $\alpha, \beta$  run over the different chiral superfields. Since we have only one chiral superfield, they only take one value. The same holds for the gauge indices $A,B$, that only take one value, namely that of the shift symmetry.
The result is
 \begin{align}
P_L \nu &=  \left(  \frac{2}{s + \bar s} - b\right) \left( \frac{1}{\sqrt 2} \frac{a e^{bs} }{s + \bar s} \chi + \frac{i}{2}c P_L \lambda \right),
\end{align}
Due to the super-BEH effect, the Golstino will give mass to the gravitino and it will disappear from the set of physical fields. As a result, the mass matrix for the fermions becomes
\begin{align}
 m = \begin{pmatrix} 
m_{\alpha \beta} + m_{\alpha \beta}^{(\nu)}
&
m_{\alpha B} + m_{\alpha B}^{(\nu)}
  \\
m_{A \beta} + m_{A \beta}^{(\nu)}
    &
   m_{AB} + m_{AB}^{(\nu)} 
      \end{pmatrix},
\end{align}
where the quantities $m_{\alpha \beta}$, $m_{\alpha A}$, $m_{AB}$ are given by
  \begin{align}
m_{\alpha \beta} &= e^{K/2} \nabla_\alpha \nabla_\beta W = \frac{a e^{bs}}{(s + \bar s)^3} \left( b^2 (s + \bar s)^2 - 2 b (s + \bar s) + 2 \right), \notag  \\
 m_{\alpha A} &= i \sqrt 2 \left( \partial_\alpha \mathcal P_A  - \frac{1}{4} f_{AB \alpha} (\text{Re}f)^{-1 \ BC} \mathcal P_C \right) =  i \sqrt{2} c \left( \frac{3}{(s + \bar s)^2} - \frac{b}{2} \frac{1}{(s + \bar s)} \right) = m_{A \alpha}, \notag \\
m_{AB} &= \frac{-1}{2} e^{K/2} f_{AB \alpha} g^{\alpha \bar \beta} \bar \nabla_{\bar \beta} \bar W =\frac{-a}{4} e^{b \bar s} (s + \bar s) \left( b - \frac{2}{s + \bar s} \right) ,\end{align}
 where $(s + \bar s)$ should be evaluated in the minimum of the potential, i.e. equations (\ref{bsbars}) and (\ref{V_minkowski}) hold.  
The corrections due to the BEH-effect are given by
\begin{align}
m_{\alpha \beta}^{(\nu)} &= \frac{-4}{3 m_{3/2}} \left( \delta_s \chi_\alpha \right) \left( \delta_s \chi_\beta \right) = \left( \frac{2}{s + \bar s} - b \right)^2 \frac{-2}{3} \frac{a}{s + \bar s} e^{\frac{b}{2} (s + \bar s)}, \\
m_{\alpha A}^{(\nu)} &= \frac{-4}{3 m_{3/2}} \left( \delta_s \chi_\alpha \right) \left( \delta_s P_R \lambda_A \right) = \left( \frac{2}{s + \bar s} - b \right)^2  \frac{-\sqrt 2}{3} ic, \\
m_{AB}^{(\nu)} &= \frac{-4}{3 m_{3/2}} \left( \delta_s P_R \lambda_A   \right) \left( \delta_s P_R \lambda_B \right) = \left( \frac{2}{s + \bar s} - b \right)^2 \frac{1}{3} \frac{c^2}{a} (s + \bar s) e^{-\frac{b}{2} (s + \bar s)},
\end{align}
where one should again evaluate $s + \bar s$ at the minimum of the potential. 
Next, due to a field redefinition in order to obtain canonically normalized kinetic terms, the 
entries in the fermion mass matrix corresponding to $m_{\alpha \bar \beta}$, $m_{\alpha A}$ and $m_{AB}$ should be multiplied 
by a factor $\alpha^2/2b^2$, $\sqrt{\alpha/2b}$ and $b/\alpha$ respectively.
As a result, the full ($2 \times 2$) mass matrix is 
\begin{align}
m =
\begin{pmatrix} \frac{\alpha^2}{2b^2} \frac{ab^3}{3 \alpha^3} e^{\alpha/2} \left( - 2 + 2 \alpha + \alpha^2 \right) & 
- \sqrt{\frac{\alpha}{2b}}  \frac{ic b^2}{3 \sqrt 2 \alpha^2} \left( -10 - 5 \alpha + 2 \alpha^2  \right)   \\ 
- \sqrt{\frac{\alpha}{2b}} \frac{ic b^2}{3 \sqrt 2 \alpha^2} \left( -10 - 5 \alpha + 2 \alpha^2  \right)     &
 \frac{b}{\alpha}  \frac{e^{-\alpha/2}\left( \alpha - 2  \right)}{12 a \alpha}  \left( -3 a^2 \alpha e^{\alpha} + 4 b c^2 \left( \alpha - 2\right) \right)        \end{pmatrix} .
 \notag 
 \end{align}
By inserting equation (\ref{V_minkowski}), the determinant of $m$ is 
\begin{align}
\text{Det}(m) &= -\frac{b^4 c^2}{6 \alpha^4 \left( -2 -4 \alpha + \alpha^2 \right)} \left( -40 - 240 \alpha - 102 \alpha^2 + 94 \alpha^3 + 15 \alpha^4 - 15\alpha^5 + 2 \alpha^6 \right)\notag \\  &= 0
\end{align}
The determinant has a root at the earlier defined value of $\alpha = -0.183268... $, so that the mass matrix has indeed a zero mode, corresponding to the reduction of the rank of the full mass matrix $m$, because the Goldstino disappears.
The (squared) mass of the fermion is thus given by the trace of the mass matrix squared Tr($m^\dagger m$) 
(or alternatively, one can square the difference between the diagonal elements in the matrix,
\begin{align}
 m_f^2 &= \text{Tr}(m^\dagger m ) \notag \\ &=  a^2 b^2 \frac{e^\alpha \left( 116 + 68 \alpha - 15 \alpha^2  - 4 \alpha^3 + 8 \alpha^4\right)}{144 \alpha^2}.
\end{align}
By using the numerical value of $\alpha$, the fermion mass can be written in terms of $m_{3/2}$
\begin{align}
m_f  &\approx 0.846 \ m_{3/2} .
\end{align}

\vfill
\end{document}